\documentclass[reprint, twocolumn]{revtex4-2}

\usepackage{graphicx}
\usepackage{chemformula}
\usepackage{lipsum}
\usepackage{graphicx}
\usepackage{subfigure}
\usepackage{color}
\usepackage{amsmath}
\usepackage[linktocpage,colorlinks=true,linkcolor=blue,citecolor=blue,urlcolor=blue,breaklinks=true]{hyperref}
\usepackage{verbatim}
\usepackage{breakcites}
\usepackage{soul}
\usepackage{bbold}
\usepackage{gensymb}
\usepackage{textcomp}
\usepackage{amssymb}
\usepackage{mathtools}\usepackage{xcolor}
\usepackage{verbatimbox}
\usepackage{multibib}
\usepackage{wasysym}
\usepackage{xcolor,colortbl}
\usepackage{enumitem}
\usepackage{array}
\usepackage{cellspace}

\def\p{{\partial}}

\begin{document}
	
\title{
	{Pure hydrodynamic instabilities in active jets of ``puller'' microalgae}
}

\author{Isabelle Eisenmann$^1$, Marco Vona$^2$, Nicolas Desprat$^1$, Takuji Ishikawa$^3$, Eric Lauga$^{2,*}$, Rapha\"el Jeanneret$^{1,*}$ } 
\affiliation{
	$^1$Laboratoire de Physique de l'\'Ecole normale sup\'erieure, ENS, Universit\'e PSL, CNRS, Sorbonne Universit\'e, Universit\'e Paris Cit\'e, F-75005 Paris, France.\\
	$^2$Department of Applied Mathematics and Theoretical Physics, University of Cambridge, Wilberforce Road, Cambridge CB3 0WA, UK.\\ $^3$Department of Biomedical Engineering, Tohoku University, 6-6-01, Aoba, Aramaki, Aoba-ku, Sendai 980-8579, Japan.}	
\email[Correspondence: ]{raphael.jeanneret@phys.ens.fr, e.lauga@damtp.cam.ac.uk}

\begin{abstract}


{Active fluids can develop spontaneous flow instabilities and complex patterns. However, spatio-temporal control of active particles 
has remained challenging, despite its relevance in biological and applied contexts. 
Here, we harnessed phototaxis to steer millions of swimming ``puller" \textit{Chlamydomonas reinhardtii} algae to create active jets and control both pearling and buckling instabilities through the preferential orientation of the cells. Our experiments, supported by a full analytical model and simulations, confirm long-standing predictions that self-generated flows can lead to jet destabilization. Our results further indicate that pullers can behave analogously to pushers when their orientation is properly tuned, and demonstrate how light enables efficient control of active fluids. }

\bigskip

%

\end{abstract}

\maketitle

Active suspensions are known to exhibit rich and diverse behaviours as a result of interactions via collectively driven flows~\cite{koch_collective_2011, ishikawa_coherent_2008}. In biological suspensions, large-scale fluid motion results from the superposition of the local dipolar flows created by the organisms~\cite{lauga_hydrodynamics_2009}. Notably, the sign of each organism-dependent dipole defines two categories of swimmers: ``pushers'', like the flagellated bacterium \textit{E.~coli}, expel fluid along their swimming direction and draw it inwards from their sides, while ``pullers'', like the microalga \textit{C.~reinhardtii}, attract fluid along their swimming direction and push it away from their sides.  

This asymmetry between pushers and pullers is known to have profound consequences on collective behavior. Suspensions of {freely swimming} pushers are unstable to orientation perturbations above a critical cell density, resulting in well-known {bacterial turbulence} patterns ~\cite{saintillan_instabilities_2008, subramanian_critical_2009}. In contrast, pullers are stable in three dimensions (3D), but can exhibit enhanced clustering in two dimensions (2D)~\cite{saintillan_instabilities_2008, bardfalvy_collective_2023}. {To our knowledge, no instabilities have yet been observed experimentally in suspensions of pullers interacting only through their self-generated flows}. {However}, recent studies of active jets have shown that when the {agents} have a preferential orientation, both pushers and pullers are prone to hydrodynamic instabilities 
~\cite{jibuti_self-focusing_2014, miles2019active, ishikawa_instability_2022}. Such jets frequently form in nature due to the interplay between the flow and an aligning stimulus such as light~\cite{williams_photo-gyrotactic_2011}, gravity~\cite{pedley_growth_1988, pedley_new_1990}, or an external magnetic field~\cite{thery_self-organisation_2020, waisbord_destabilization_nodate}.  
Inside {a jet}, cells swimming behind each other effectively pull or push on their neighbors depending on their pusher/puller nature. A jet of pullers is thus prone to {pearling}~\cite{lauga_clustering_2016}, while a jet of pushers is prone to buckling~\cite{lauga_zigzag_2021}. Such instabilities have indeed been observed in both 2D and 3D simulations~\cite{ishikawa_instability_2022}, and have been the subject of theoretical investigation in one dimension \cite{lauga_clustering_2016,lauga_zigzag_2021}. However, an experimental realization of both instabilities is still lacking ({although} a related study reported {buckling} of bacteria in liquid crystals~\cite{turiv_polar_2020}), in part because strongly aligning a dense cell suspension without disturbing the fluid is experimentally challenging. Here, using phototactic microalgae {steered by light}, we reproduce both the {pearling} and the zigzag instabilities in a dense puller jet{,} {showing that pullers and pushers can be equivalent when the cells' orientation is properly tuned. Importantly, light enables {us to precisely control not only the time-dependent cell swimming direction}, but also the magnitude of {cell} alignment by increasing the intensity}. We compare our experimental results {with a new continuum theory}, presented in detail in a joint article~\cite{joint_article}, and with simulations of microswimmers suspensions, demonstrating that these instabilities are indeed purely hydrodynamic in nature. 
Taken together, these results show how an assembly of aligned pullers can be destabilized purely by their own flows.

\begin{figure} 
	\includegraphics[width=\columnwidth]{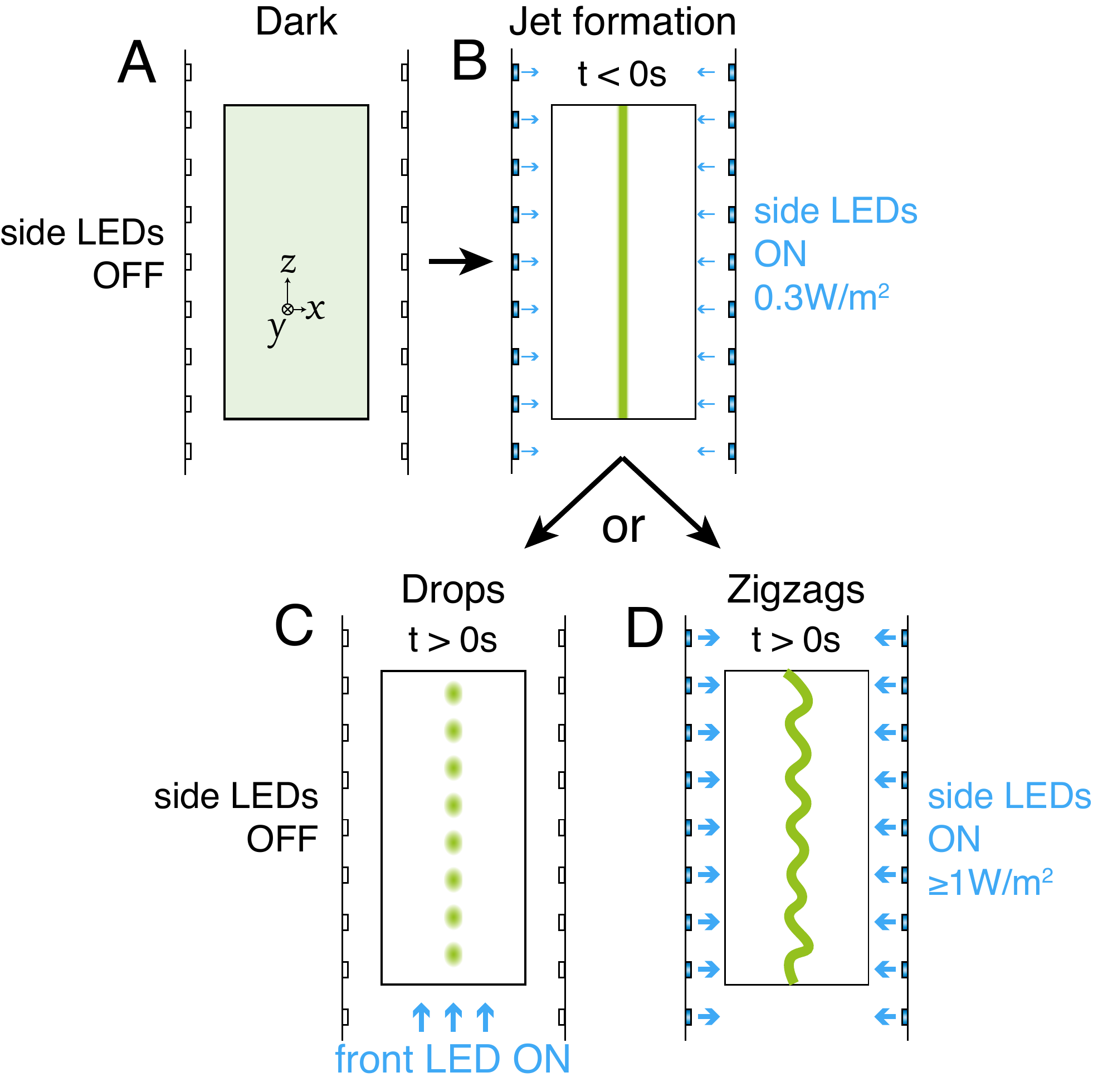} 
	\caption{
		Schematics of the experimental setup (top view).
		\textbf{A}: Initially the cell suspension is homogeneous.
		\textbf{B}: Side LED arrays are switched on at a weak intensity to form a dense band of cells at the center of the dish.
		\textbf{C}: When a collimated LED aligned with the band is switched on (side LED arrays off), the jet destabilizes into {clusters}.
		\textbf{D}: If instead the side light intensity is higher than $\rm 1W/m^2$ (without the collimated LED), the band buckles into a zigzag structure.
		\label{fig:setup}
	}
\end{figure}

The model unicellular alga \textit{C.~reinhardtii} (diameter, $\rm 10\ \mu m$) {self-propels} via the synchronous beating of its two front-mounted cilia~\cite{jeanneret_brief_2016}. The strain CC125 was grown in Tris-Acetate-Phosphate medium under light/dark cycles (16h/8h). Cells were harvested in the exponential phase and diluted to reach $\sim10^5$ cells/mL, then acclimated twenty minutes in the dark with aeration. In these conditions,  cells only performed negative phototaxis (i.e.~fled light). We loaded $3$mL of the suspension into a rectangular dish  (${\rm W\times H\times L}=30\times13\times73$mm), reaching a liquid height of $\sim1$mm. The dish was placed between two 9cm-long parallel arrays of white LEDs (PowerLED 12V, 6000-7000K). After a few minutes of weak illumination ($\rm 0.3\ \text{Wm}^{-2}$), almost all cells were found in the middle of the dish, in a dense horizontal band spanning its whole length (Fig.~\ref{fig:setup}A-B). As the LEDs were slightly higher than the liquid surface, the band was located at the bottom of the chamber. The cell density under similar illumination geometries was previously measured to be $\rho\sim3\times10^9$cells/mL~\cite{eisenmann_light-induced_2024}. All movies were recorded using a far-red visualization light to which cells are blind (Thorlabs M730L5, 730nm). 

\begin{figure}
	\includegraphics[width=\columnwidth]{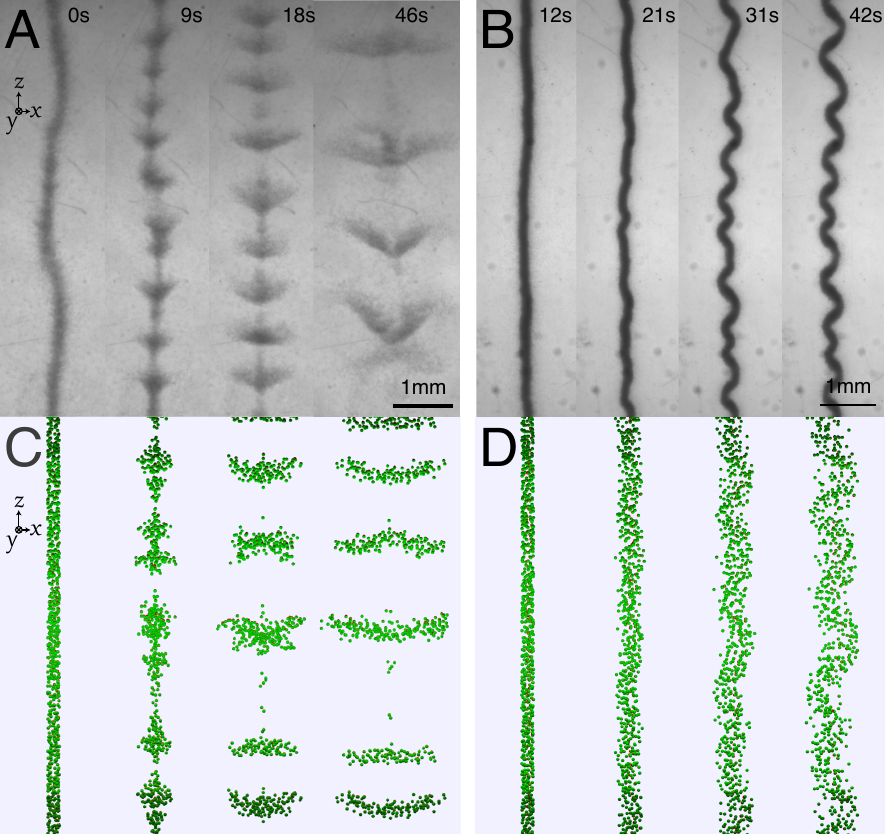} 
	\caption{Instabilities in active jets in experiments (top) and simulations (bottom). 
		\textbf{A:}  {Pearling} instability in experiments (top view, light coming from the bottom of  image).
		\textbf{B:} Zigzag instability in experiments  (top view, light coming from both sides).
		\textbf{C and D:} Simulations of pullers {oriented} parallel (panel C) or perpendicular (panel D) to the jet.  Details of the simulations can be found in \cite{SM,joint_article}.		
		Initial cell density and dimensionless stresslet are respectively $\phi_0=0.07$ and $\beta=1$, {respectively}. Snapshots taken at times $0, 66, 132, 198 t^*$ with $t^*=\mu  {a_s}^3/(S\phi_0)$.
		\label{fig:imgs}
	}
\end{figure}

We {observed} that two instabilities can arise in such a dense band, depending on the  orientations of the cells. In the first set of experiments, cells were constrained to swim parallel to the band by switching off the side LED arrays and switching on a front collimated LED (Thorlabs MCWHL8) at time $t=0$ (Fig.~\ref{fig:setup}C). The band was immediately set into motion, migrating in the direction opposite to the light. After a few seconds, the jet started to break into {clusters}  migrating at $\rm \sim 100\ \mu m\ \text{s}^{-1}$, close to the speed of individual algae (Fig.~\ref{fig:imgs}A and Movie S1 \cite{SM}). The regularity of the spacing between {clusters} suggested the existence of an instability with a well-defined wavelength, which appeared directly related to the initial band width. After the initial breakup, {clusters} were seen to stretch laterally, before a secondary instability 
deformed them into a stable ``V'' shape with large vortices of alternating directions (Movie S2 and S3 \cite{SM}).

In the second set of experiments, we simply increased the intensity of the two parallel LED arrays at $t=0$, without the collimated LED (Fig.~\ref{fig:setup}D). 
In this setup, the  main swimming direction of the cells was therefore perpendicular to the band. The sudden increase in light intensity first triggered a transient positive phototactic response of the algae~\cite{uhl_adaptation_1990,takahashi_photosynthesis_1993}, causing the band to briefly inflate.  After a few seconds, cells {again} proceeded to flee the lights and a regular  {zigzag} structure similar to~\cite{katzmeier_emergence_2022, junot_large_2023, shoham_hamiltonian_2023, turiv_polar_2020} appeared, whose amplitude rapidly stabilized (Fig.~\ref{fig:imgs}B and Movie S4 \cite{SM}). Once again, this appeared to be the result of an instability, characterised by a well-defined wavelength, directly related to the {jet} width. The final zigzag structure exhibited no net motion but was still highly dynamical, with macroscopic cell recirculation and merging and splitting events of the peaks (Movie S5 \cite{SM}).

\begin{figure} 
	\includegraphics[width=\columnwidth]{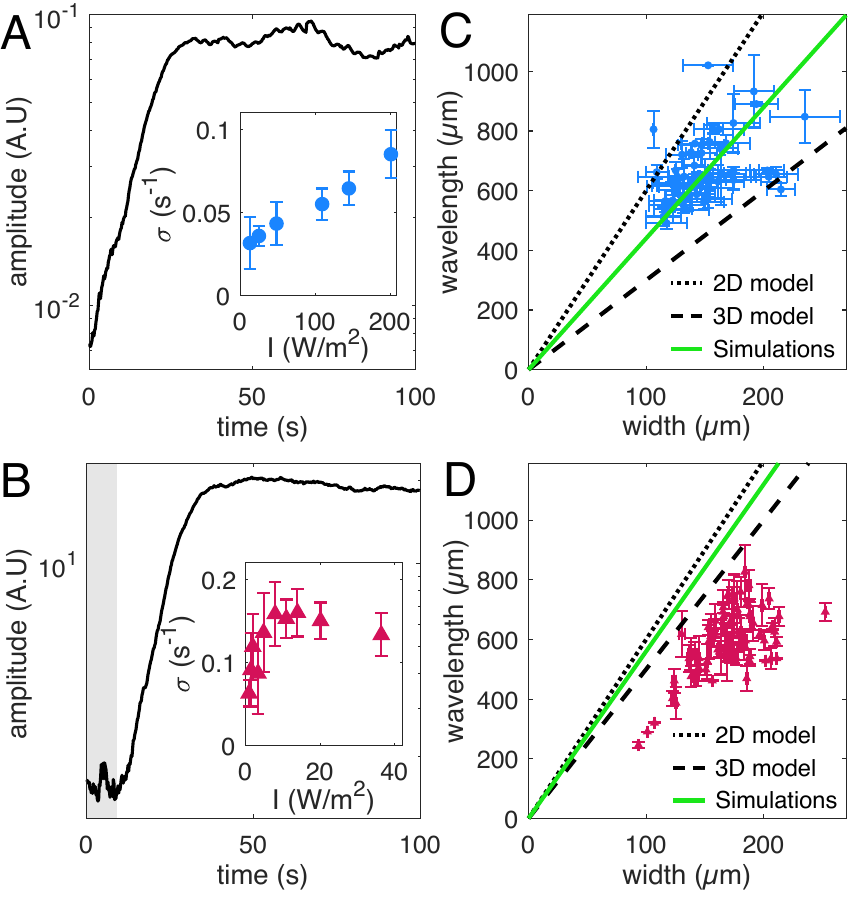} 
	\caption{Left:  
		Time evolution of the perturbation amplitude for {clusters} (\textbf{A}) and zigzags ({\textbf{B}}); inset:  growth rate at multiple light intensities. {Error bars are STD of 3 to 5 replicates}. Right: Selected wavelength vs.~initial jet width for {clusters} ({\textbf{C}}) and zigzags (\textbf{D}); a linear fit of the experimental data gives slopes $\lambda^{\text{ {pearling}}}_{\text{exp}}\sim4.4w$ and $\lambda^{\text{zigzag}}_{\text{exp}}\sim3.5w$. {Each point represent a single experiment ; see \cite{SM} for details on the measurements of amplitude, wavelength and width.}
		\label{fig:drops}
	}
\end{figure}

To characterize the observed instabilities, we measured the evolution of their amplitudes and wavelengths over time (see Supplementary Material~\cite{SM} for details).  
The growth rate of both instabilities increased with light intensity, while the jet did not exhibit any discernible instabilities at low intensities. A closer comparison revealed that the  {pearling} instability required more light than the zigzags (Fig.~\ref{fig:drops}A, B): while for zigzags the growth rate saturated at  $\sigma^{\text{zigzag}}_{\text{exp}}=0.15\ \text{s}^{-1}$ above $10~{\rm W/m^2}$, for {pearling} we only reached {$  {\sigma}^{\text{ {pearling}}}_{\text{exp}}=0.08\ \text{s}^{-1}$} at $200 ~{\rm W/m^2}$. 
Similarly, in the  pearling experiments the {approximate} threshold intensity required to destabilize the jet was higher than in the zigzag experiments {($\lesssim 5\ {\rm W/m^{2}}$ vs.~$\lesssim \rm 1\ {\rm W/m^{2}}$)}.  
{Below these values, the jet remained {stable} in the zigzag setup, while it laterally diffused before {any observable} destabilization in the {pearling} setup.}

We proceeded to measure the wavelengths of the instabilities, defined either as the average distance between {clusters}  or between the zigzag{'s} peaks, in both cases at the end of the exponential growth regime. We obtained a linear correlation {between} the selected wavelength {and} the jet{'s} width, regardless of the light intensity used. For the {clusters}, the selected wavelength was about $4$ times the initial jet width, as measured between $\rm 0<t<3\ \text{s}$ (Fig.~\ref{fig:drops}C); {on the other hand,} for the zigzags  the wavelength was closer to $3$ times the initial band width (Fig.~\ref{fig:drops}D).

The large degree of similarity between these experiments and previous active suspensions models~\cite{jibuti_self-focusing_2014, lauga_clustering_2016, lauga_zigzag_2021, ishikawa_instability_2022} suggests that the observed instabilities are hydrodynamic in nature. To confirm this hypothesis, we compared our experimental measurements with a bespoke analytical model and with direct hydrodynamic simulations. The details of the model, including the derivation of the hydrodynamic equations for the evolution of the jet in both 2D and 3D, may be found in {the End Matter and in} ~\cite{joint_article}. 
{In short, we derived a continuum model for an active suspension of aligned particles (with fixed swimming direction) initially arranged in a coherent jet, either three-dimensional (cylindrical) or two-dimensional (sheet-like), surrounded by cell-free fluid. We obtained effective equations for the bulk flow, boundary conditions at the jet edge, and a conservation law for the swimmer volume fraction. The model predicts analytically that jets are unstable to certain shape perturbations, extending prior one-dimensional analyses~\cite{lauga_zigzag_2021, lauga_clustering_2016} to bulk flows. The main results are summarized next.}

For pullers ($S>0$), the largest growth rate  
 occurs for the {axisymmetric mode}, meaning that such {a} jet is unstable to pearling. 
 The axial wavenumber maximizing {the growth rate} $\sigma$ gives the selected wavelength. We {thus} retrieve the linear dependence of the wavelength on the initial jet width $w$ observed in experiments: $\lambda_{3D}^{\text{ {pearling}}}=2\pi/k\sim 3w$.
A similar analysis of a 2D jet reveals an analogous pearling instability, with a higher $\lambda_{2D}^{\text{ {pearling}}}=2\pi w$~\cite{joint_article}. These analytical predictions are in good agreement with the experimental  {pearling} {wavelength} $\lambda^{\text{ {pearling}}}_{\text{exp}}\sim 4w$, lying between the 2D and 3D theory (Fig.~\ref{fig:drops}C), {consistently} with the fact that, in the experiments, the cells swim close to the bottom surface of the dish, making the jet appear like a flattened cylinder.

Conversely, for pushers ($S<0$) {the most unstable} mode is {helical}. A pusher jet is thus prone to  {zigzags}, with an estimated $\lambda_{3D}^{\text{zigzag}}\sim 5w$ in 3D. A similar analysis in a 2D setup also unveils a  {zigzag} instability, this time with $\lambda_{2D}^{\text{zigzag}}=2\pi w$. Despite the experimental jet not consisting of pushers but of rotated pullers, these results are consistent with the experimental wavelength (Fig.~\ref{fig:drops}D). This can be understood physically from the microscopic flows driving the instabilities: pullers generate flows which are very similar to pushers after a 90$\degree$ rotation~\cite{lauga_hydrodynamics_2009, drescher_direct_2010, drescher_fluid_2011}. A line of pullers swimming side by side (like our zigzags experiments) is thus mathematically similar to a line of pushers swimming in a single file (like in the model). A second important difference with the theory is that, in the zigzag experiments, there is no net motion of the jet due to swimming. At the center of the  dense band it is probably difficult for cells to align with the light due to shading effects and steric collisions \cite{eisenmann_light-induced_2024}. However{,} for an approximately uniform orientation distribution within the jet, the bulk active stresses should also approximately cancel and thus, as in the model, the dynamics should be driven by the boundary stresses. In experiments, cells are indeed aligned almost perpendicularly to the jet close to the interfaces. Nevertheless, these differences may explain the slight discrepancy between the experimental wavelength ($\lambda^{\text{zigzag}}_{\text{exp}}\sim 3w$) and the higher analytical values. 

{
Another testable prediction of the model is the growth rate itself. Our theoretical analysis yields the {following}  predictions in the dilute limit ($\phi_0\ll1$)}{:}
%
\begin{subequations}
\begin{align}
&\sigma^{\text{ {pearling}}}_{\text{3D}}\sim 0.097\times\frac{|S|\phi_0}{\mu V_s} \label{eq:sdrop},\\
&\sigma^{\text{zigzag}}_{3D}\sim 0.024\times\frac{|S|\phi_0}{\mu V_s},\label{eq:szigzag}\\
&\sigma^{\text{ {pearling}}}_{2D}=\sigma^{\text{zigzag}}_{2D}\sim 0.092 \times\frac{|S|\phi_0}{\mu V_s}. \label{sigma 2d}
\end{align}
\end{subequations}
{
In the experiments, we estimate $V_s\sim 180 \ \mu \text{m}^3$ \cite{eisenmann_light-induced_2024}  
and {$S\sim 8\times 10^{-6} \text{ kg.} \mu\text{m}^2\text{.s}^{-2}$~\cite{klindt2015flagellar}}. Taking $\rm \mu = 10^{-9}\ kg.\mu m^{-1}.s^{-1}$ {for} water, as well as $\phi_0\sim 0.2$ for the  {clusters} and $\phi_0\sim 0.5$ for the zigzags,
 {we find $\sigma^{\text{ {pearling}}}_{3D}\sim 0.9\ \text{s}^{-1}$} and {$\sigma^{\text{zigzag}}_{3D}\sim 0.5\ \text{s}^{-1}$}, while {$\sigma^{\text{ {pearling}}}_{2D}\sim 0.8\ \text{s}^{-1}$}, {$\sigma^{\text{zigzags}}_{2D}\sim 2 \ \text{s}^{-1}$}. 
 {The experimental measurements performed in a dense jet ($\sigma_{\text{exp}}\sim 0.1\ \text{s}^{-1}$) are lower than these dilute estimates, as expected from the extra resistance added by particle-particle interactions. }
{Indeed, in a dense suspension } the linear response theory is expected to provide at best an upper bound for the growth rate, as confirmed by numerical simulations (see below).
}

 We next compare experiments and theory with computations of dilute ($\phi_0=0.07$) jets and bands of pullers. These numerical simulations allow us to precisely track the onset of the instability and the long-term evolution of the jet. {In addition, in these simulations our {active particles} have {vanishing} swimming velocity but a non-zero stresslet, unambiguously showing that the instabilities arise solely from hydrodynamic interactions.} Once again, all the details can be found in our joint article~\cite{joint_article}. In short, we used the Stokesian dynamics method from~\cite{ishikawa_instability_2022}, which accurately accounts for near-field interactions {as well}. 
 Cell were subjected to a torque 
 maintaining cells aligned with the light's main direction. Despite phototaxis not involving an external torque, we demonstrate in ~\cite{joint_article} that the leading-order dynamics are unchanged in the dilute limit.

 In the first batch of simulations, wherein the aligning torque kept pullers aligned with the  axis of the jet, the jet indeed broke up into {clusters}. Strikingly, the instability development remained very similar to experiments even long after the emergence of the instability: soon after their formation, {clusters} stretched laterally, then slowly coalesced while taking a shape similar to the experiments (Fig.~\ref{fig:imgs}C and Movie S6 \cite{SM}). We also found a linear relationship between wavelength and {jet} width $\lambda^{\text{ {pearling}}}_{\text{sims}}\sim 4.4w$ with the same slope as the experiments{,} again in good agreement with the theory $\lambda^{\text{ {pearling}}}_{3D}\sim3w$ and $\lambda^{\text{ {pearling}}}_{2D}=2\pi w$ (Fig.~\ref{fig:drops}C).
Furthermore, the growth {rate} appeared linearly related to $S$ and $\phi_0$ as in the model.  {An early-times exponential fit {yielded}} the estimate $\sigma^{\text{{pearling}}}_{\text{sims}}\sim  {0.121}\times |S|\phi_0/\mu V_s$, which compares favorably with the analytical prediction \eqref{sigma 2d}. 
 
When pullers were instead oriented perpendicularly to the  axis of the jet, we observed  {zigzags} with $\lambda^{\text{zigzag}}_{\text{sims}}\sim 5.6w$,  consistent  with (but slightly higher than)  experiments (Fig.~\ref{fig:drops}D and Movie S7 \cite{SM}) {and agreeing {closely} with the theory ($\lambda^{\text{zigzag}}_{3D}\sim5w$ and $\lambda^{\text{zigzag}}_{2D}=2\pi w$). Here the growth rate again appeared to evolve} linearly on $S$ and $\phi_0$ {and through an exponential fit} we {obtained} an estimate $\sigma^{\text{zizgags}}_{\text{sims}}\sim  {0.079}\times |S|\phi_0/\mu V_s$, similarly to \eqref{sigma 2d}.

 {Finally, increasing the concentration to a non-dilute volume fraction ($\phi_0 =0.12$) in these simulations confirmed that our dilute theory overestimates the growth rate at high densities.}

\begin{figure} 
	\includegraphics[width=\columnwidth]{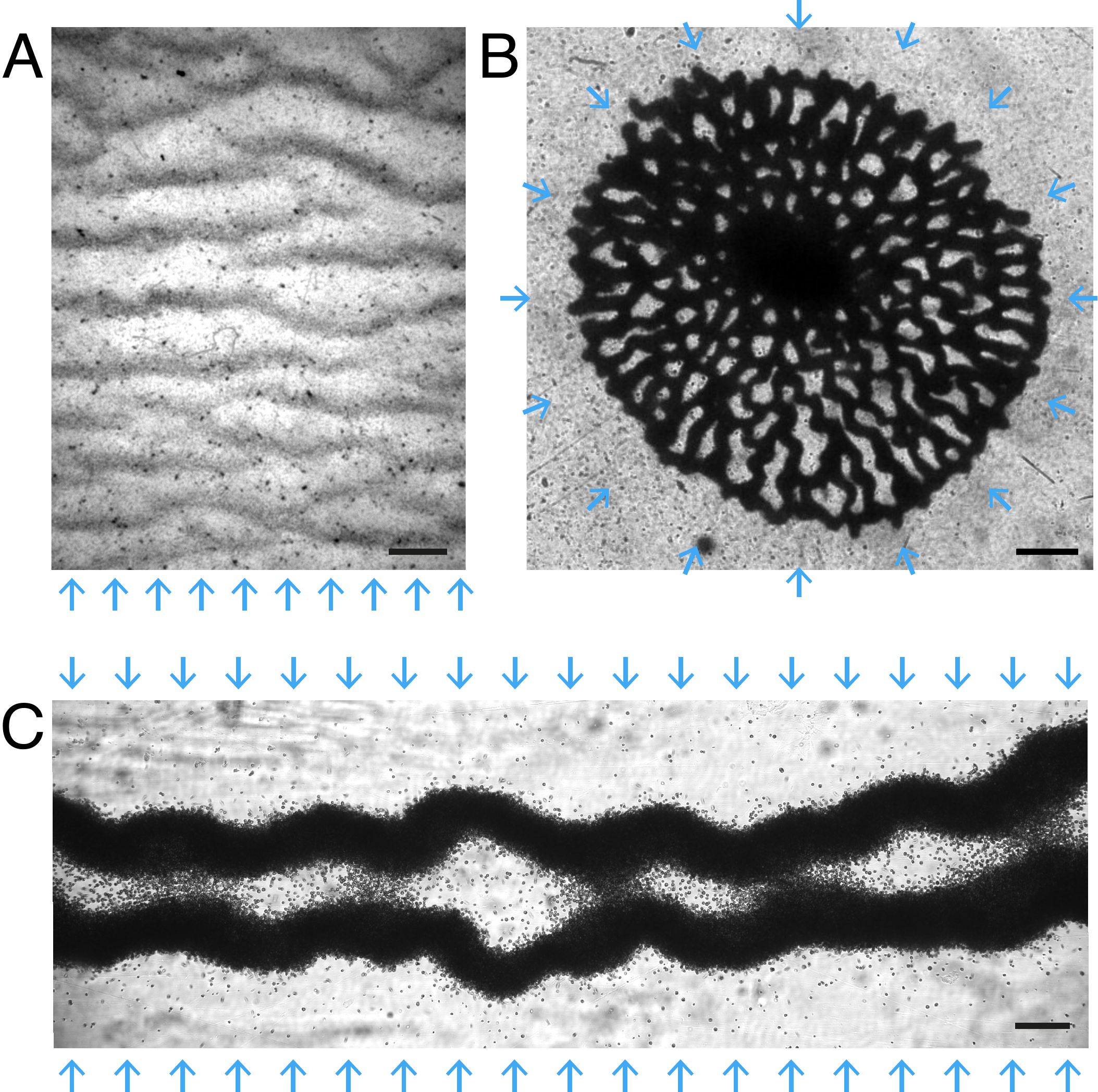} 
	\caption{{
		\textbf{A}: Travelling density bands in a 2D suspension of aligned microalgae, with light coming from an array of LED on the side (top view). Scale bar 1 mm.
		\textbf{B}: {``Bubble''}  formation in a dense suspension of cells in a circular geometry, with 16 LEDs surrounding it (top view). Scale bar {1 mm}.
		\textbf{C}: Fission of a wide band of cells in the geometry of Fig.~\ref{fig:setup}C. Scale bar {$200\ \mu \text{m}$}.}}
		\label{fig:patterns}
\end{figure}

{
{
Beyond the fundamental instabilities displayed in Fig.~\ref{fig:imgs},     hydrodynamic interactions can produce a range of patterns when the suspension{'s} geometry or external field is altered} (Fig.~\ref{fig:patterns}). For instance, {if the rectangular chamber was simply illuminated with the collimated LED (Fig.~\ref{fig:setup}C)  without first forming a jet},  {the suspension rapidly destabilized} into travelling density bands  {moving against}  the light (Fig.~\ref{fig:patterns}A), reminiscent of  bands  in simulations of 2D puller sheets under gravitactic torques~\cite{ishikawa_coherent_2008}.  Strikingly, the bands {exhibited} a secondary {``V''}-shaped deformation, visually similar to {that in} {2D} clusters and zigzags. 
{For a dense suspension with a circular geometry}~\cite{eisenmann_light-induced_2024}, 
a sudden {LED intensity increase} also {triggered} an instability {akin to} zigzags, {where a}  dense blob  {evolved} into a transient bubble-like pattern  (Fig.~\ref{fig:patterns}B), with wavy borders {resembling} the buckled bands. Finally, when the bands in Fig.~\ref{fig:setup}D {reached a sufficient initial width} {($w\gtrapprox200 {\rm \mu m}$),}  
{were seen to split in half before} before {undulating} (Fig.~\ref{fig:patterns}C), setting a limit {of $\sim 200\ \mu \text{m}$} to the maximal width at which a band buckles. This   splitting {likely} involves both hydrodynamics and shading effects~\cite{eisenmann_light-induced_2024}.
}

Our results strongly suggest that the experimentally observed pearling and zigzag instabilities are driven solely by hydrodynamic interactions, and {we established} that the theory can quantitatively capture both their  {their short-term dynamics}.
We showed that pullers and pushers can exhibit qualitatively similar behaviors in the right setting due to the similarities between their flow fields. Phototaxis proved to be a powerful tool to remotely control suspensions of swimming micro-organisms without applying forces {to} the fluid. Experimentally, designing optical landscapes is straightforward, and provides a way to control even dense suspensions at both high temporal and spatial resolutions, as well as to apply strong effective torques {to the cells (necessary in our experiments)}. {Remarkably, the magnitude of this effective torque can also be tuned by increasing the light intensity \cite{joint_article}}. {Extending previous one-dimensional approaches \cite{lauga_clustering_2016,lauga_zigzag_2021}, our analytical framework quantitatively captured 
the experimental instabilities. }
As expected, our linearized theory  overestimate{d} the growth rate compared to the dense experiments, but {could} still capture {the key dynamics of the instabilities}. {This 
model is presented in detail in our joint article~\cite{joint_article}, where}  
we derive the instability growth rates in $3$D and $2$D, and asymptotically determine the stability of individual Fourier modes. We also investigate the long-term dynamics and interactions of the puller clusters, deriving exact and self-similar asymptotic solutions in both $3$D and $2$D, {and obtaining exact solutions for the observed ``V''  instability}. We finally describe the large-time evolution of the zigzags through a slender-body theory, recovering key experimental features like wavelength coarsening. Comparisons of flow fields, growth rates and long-term dynamics reveals excellent agreement with both experiments and numerics. Our results should hold generically at low Reynolds numbers, provided {that} the active flows remain dipolar, {consistently} with extensive studies on driven colloids in alternating electric fields and, more recently, magnetic particles in oscillating fields ~\cite{isambert_electrohydrodynamic_1997, katzmeier_emergence_2022, junot_large_2023, shoham_hamiltonian_2023}. By unifying this broad class of systems under a single framework, our {work} opens new avenues for {the} precise experimental control {of active suspensions}.


%

\section*{acknowledgments}
 {This work was partly funded by EPSRC (scholarship to MV) and supported through the Junior Research Chair Programme (ANR-10-LABX-0010/ANR-10-IDEX-0001-02 PSL; R.J.) and an ED-PIF doctoral fellowship (I.E.). TI is supported by JSPS KAKENHI (No. 21H04999 and 21H05308).}

\clearpage

\section*{End Matter}

We {herein} summarise here the derivation of the model and outline the stability calculations performed for different jet geometries, the analysis of which is presented in detail in the companion article~\cite{joint_article}. We also comment on the {duality between pushers and pullers in specific setups}.

\subsection*{{Jet geometry and modelling assumptions}}

We consider a dilute suspension of force-free, identical spherical swimmers in an unbounded fluid. The swimmers are assumed to be subjected to an aligning stimulus (such as light, gravity,  {magnetic fields}, or chemical  {gradients}), with preferred direction $\mathbf e_z$. We let the swimmer radius be $a_s$, the swimmer volume be $V_s$ and we denote by $U_s$ the (constant) swimming speed of an isolated swimmer. Finally, the swimmer volume fraction {is denoted by} by $\phi(\mathbf x,t)\ll 1$. We assume that all swimmers can initially be found inside a specified region of space, which corresponds to the base state of the ``jet''.

Two different geometries will be considered, motivated by experiments:
\begin{enumerate}
    \item Cylindrical  {jet}s: all swimmers are initially positioned in the domain given in cylindrical coordinates by $0\leq r\leq a$, $-\infty<z<\infty$.\label{Cylindrical Case} 
    {(Fig.~2 of Ref.~\cite{joint_article})}
    \item Quasi two-dimensional  {jet}s: all swimmers are initially positioned in {a} sheet of finite thickness given by $-a\leq x\leq a$, $-\infty<y,z<\infty$ {in Cartesian coordinates}. The sheet is taken to be homogeneous in the $y$ direction, so that the dynamics is effectively restricted to the $xz$ plane. {(Fig.~4 of Ref.~\cite{joint_article})}
    \label{Planar case}
\end{enumerate}
In each case, the volume fraction $\phi_0$ within the initial jet is assumed constant. In order to model a jet with a sharp boundary, two key assumptions are made:
\begin{enumerate}
    \item Swimmers re-orient rapidly on the timescale of the flow, so that their swimming direction is quasi-steadily $\mathbf e_z$.
    \item Spatial diffusion is negligible, so that swimmers do not leave the  {jet} as a result of stochastic effects.
\end{enumerate}
The swimmers drive a large-scale flow $\mathbf u$ by each exerting a stresslet $S(\mathbf e_z\mathbf e_z-\mathbf I/3)$ on the fluid, where the strength is $S>0$ for a puller swimmer and $S<0$ for a pusher. {For \textit{C.~reinhardtii}, $S$ fluctuates during the beat cycle, but its {period}--averaged value is positive \cite{klindt2015flagellar}. We will later see that such {period}--average can be viewed as an ``effective'' strength in our theory, so that \textit{C.~reinhardtii} behaves as a puller in our system}. In a dilute suspension, $S$ is taken to be the same as for a swimmer in isolation, up to $\mathcal O(\phi)$. Similarly, any aligning torque must scale with the $\mathcal O(\phi)$ flow and can thus be neglected at leading order.

\subsection*{Derivation of the main equations}

A jet formally partitions space into a swimmer-laden region $\mathcal P^-$ and a swimmer-free region $\mathcal P^+$, with common boundary $\p \mathcal P$. The first equation governing the jet enforces swimmer conservation: swimmers are advected along the streamlines of the incompressible field $\mathbf u+U_s\mathbf e_z$, so that the volume fraction is constant along these characteristic curves, and thus
\begin{align}
 &\frac{\mathrm d}{\mathrm dt}\phi[\mathbf x(t),t]=0, & &\dot{\mathbf x}(t)=\mathbf u(\mathbf x,t)+U_s \mathbf e_z. \label{Swimmer Conservation}  
\end{align}
The solution of Eq.~\eqref{Swimmer Conservation} consists of two time-dependent regions $\mathcal P^-(t)$ and $\mathcal P^+(t)$, such that $\phi\equiv \phi_0$ in $\mathcal P^-$, $\phi\equiv 0$ in $\mathcal P^+$ and points on the boundary $\p \mathcal P$ are advected along $\mathbf u(\mathbf x,t)+U_s \mathbf e_z$.

Secondly, the flow $\mathbf u$ and pressure field $q$ driven by the swimmers may be found at order $\mathcal O(\phi)$ by superposing their local dipolar flows{:}
\begin{equation}
\nabla q(\mathbf x)-\mu\nabla^2\mathbf u(\mathbf x)=-\frac{S}{V_s}\mathbf e_z\mathbf e_z\cdot\int_{\mathbb R^3} \phi(\mathbf x_0)\nabla_0\delta^{(3)}(\mathbf x-\mathbf x_0)\mathrm d^3\mathbf x_0.    
\end{equation}

Splitting up the integral over {$\mathcal P^-$, $\mathcal P^+$} and applying the divergence theorem we find (for $\mathbf x\not\in \p\mathcal P$)
\begin{align}
\nabla q(\mathbf x)-\mu\nabla^2\mathbf u(\mathbf x)=\frac{S}{V_s}\oint_{\p \mathcal P}\delta^{(3)}(\mathbf x-\mathbf x_0)[\phi]_-^+\mathbf e_z(\mathbf e_z\cdot\mathbf n)\mathrm dA \label{Forced Stokes P}  ,
\end{align}
where $\mathbf n$ is the boundary unit normal pointing into $\mathcal P^+$ and $[\phi]^+_-$ denotes the concentration jump across $\p\mathcal P$. The solution of Eq.~\eqref{Forced Stokes P} may be found by linear superposition, and it is given by
\begin{equation}
\mathbf u(\mathbf x)=\frac{S}{V_s}\oint_{\p \mathcal P}\mathbf J(\mathbf x-\mathbf y)\cdot\mathbf e_z(\mathbf e_z\cdot\mathbf n)[\phi]_-^+\mathrm dA.\label{u solution}   
\end{equation}
where $\mathbf J$ is the Oseen tensor ~\cite{happel1983low, kim2013microhydrodynamics}. {From Eq.~\eqref{u solution}, } the flow $\mathbf u$ and pressure $q$ are the unique solution of {the following} Stokes problem forced on $\p \mathcal P${:}
\begin{subequations}
\begin{align}
&\nabla q-\mu\nabla^2 \mathbf u=\mathbf 0, \qquad\mathbf x\in \mathcal P^-(t),\ \mathcal P^+(t),\\
&{\nabla \cdot \mathbf u=0, \qquad\mathbf x\in \mathcal P^-(t),\ \mathcal P^+(t),}\\
&{[\mathbf u]^+_-=\mathbf 0,}\\
&\left[-q\mathbf I+\mu(\nabla\mathbf u+\nabla\mathbf u^{\text T})\right]^+_-\cdot \mathbf n=-\frac{S}{V_s}[\phi]_-^+\mathbf e_z(\mathbf e_z\cdot\mathbf n).\label{Stress continuity condition}
\end{align}
\label{Stokes problem to solve}
\end{subequations}
Eqs.~\eqref{Swimmer Conservation}{, }\eqref{Stokes problem to solve} completely determine the shape evolution of the initial jet, in a way that we now summarise in the two geometries of interest.

\subsection*{Stability analysis}

In the case \#\ref{Cylindrical Case} of a three-dimensional, cylindrical jet, the solution of 
Eqs.~\eqref{Swimmer Conservation}{, }\eqref{Stokes problem to solve} has $\mathbf u=\mathbf 0$ in the base state, as there is no active force component normal to the jet's boundary. The stability of the jet is then investigated by perturbing the radius to $\displaystyle r=a+\varepsilon \eta=a(1+\varepsilon e^{st+\mathrm ikz+\mathrm in\theta})$ and solving for the growth rate {$\sigma=\Re(s)$}. This is done by expanding all field variables with respect to $\varepsilon$ and solving Eqs.~\eqref{Swimmer Conservation}{, }\eqref{Stokes problem to solve} for the velocity and pressure fields at order $\mathcal O(\varepsilon)$. This resulting expressions depend on six unknown coefficients. By imposing the stress boundary condition, Eq.~\eqref{Stress continuity condition}, and the advection equation for the jet's boundary
\begin{align}
\left[\frac{\p}{\p t}+(\mathbf u+U_s\mathbf e_z)\cdot\nabla\right](r-a-\varepsilon \eta)&=0,  & &\mathbf x\in\p \mathcal P,  
\end{align}
a homogeneous linear system for the coefficients is obtained. Requiring that the determinant must vanish in order for the flow to be non-zero uniquely determines the growth rate ${\sigma}$ as
\begin{widetext}
\begin{align}
  {\sigma}=\frac{S\phi_0}{2\mu V_s}\frac{I_n(\xi) \left[K_n(\xi) \left(2n^2 +\xi^2 \right)-n\xi K_{n+1} (\xi)\right]-I_{n+1} (\xi)\left[\xi^2 K_{n+1} (\xi)-n\xi K_n (\xi)\right]}{\xi I_n (\xi)K_{n+1} (\xi)+\xi K_n (\xi)I_{n+1} (\xi)}. \label{Most General Growth Rate Appendix} 
\end{align}
\end{widetext}
Here, $\xi=ak$ and the $I_n$, $K_n$ are modified Bessel functions of the first and second kind. From Eq.~\eqref{Most General Growth Rate Appendix}, we find that the largest growth rate for a puller jet ($S>0$) occurs for $n=0$ (axisymmetric pearling mode) and $\lambda^*\sim 5.8447\times a${,} and is given by ${\sigma^*}\sim 0.0968\times S\phi_0/\mu V_s$. For a pusher jet ($S<0$), we conversely find that the largest growth rate has $n=1$ (helical mode) and $\lambda^*\sim 9.8943\times a${,} and is given by ${\sigma^*}\sim 0.0243 \times |S|\phi_0/\mu V_s$. We further show in~\cite{joint_article} that Eq.~\eqref{Most General Growth Rate Appendix} also holds for a {rapidly-varying} time-dependent stresslet $S(t)$, if $S$ is replaced by the {period}-average $\langle S\rangle$. This holds provided that $S$ is spatially uniform and the corresponding Stokes number is not too large. This makes our results applicable to swimmers like \textit{C.~reinhardtii}, which alternates between a pusher and puller state during its beat pattern~\cite{klindt2015flagellar}.

A similar analysis can be performed for the sheet-like jet from case \#\ref{Planar case}. This time, we perturb the boundary so that the subsequent jet has $X^-(z,t)<x<X^+(z,t)$, $-\infty<y,z<\infty$. We consider two forms of the perturbation:
\begin{enumerate}
    \item Sinuous (i.e.~in phase): $X^+=a\left(1+\varepsilon e^{st+\mathrm ikz}\right)$, $X^-=a\left(-1+\varepsilon e^{st+\mathrm ikz}\right)$;
    \item Varicose ({i.e.}~antiphase): $X^+=a\left(1+\varepsilon e^{st+\mathrm ikz}\right)$, $X^-=a\left(-1-\varepsilon e^{st+\mathrm ikz}\right)$.
\end{enumerate}
In each case, the growth rate ${\sigma}$ is found analogously to case \#\ref{Cylindrical Case}, expanding all field variable with respect to $\varepsilon$ and solving for the velocity field at order $\mathcal O(\varepsilon)$. For a sinuous/varicose perturbation, the growth rates found from the zero-determinant condition are 
 \begin{align}
{\sigma}&=\mp\frac{S\phi_0}{2\mu V_s}\xi e^{-2\xi}.\label{2D Growth Rate Varicous} 
    \end{align}
  {respectively.} Therefore, a sinuous perturbation is unstable when $S<0$ (pusher jet) and varicose one when $S>0$ (puller jet). These modes are the two-dimensional equivalents of the helical/pearling instabilities in the three-dimensional case. {The} largest growth rates and corresponding wavelengths, identical for both perturbations, are found to be $\lambda^*\sim 12.5664\times a$ and ${\sigma^*}\sim 0.0920\times |S|\phi_0/\mu V$.

{In the joint article \cite{joint_article}, we further characterise the ranges of $\xi$ and $n$ which make the cylindrical jet unstable for pushers and pullers. We then analyse the late-time evolution of the pearling instability, deriving exact and self-similar asymptotic solution for the expansion of the clusters. We further obtain an exact solution for the evolution of a tilted cluster, motivated by the experimental observation of the ``V''-instability. Moreover, we propose an adapted slender-body theory to analyse the late-time shape evolution of the $3$-dimensional zigzags. We recover the slow timescale for shape evolution and the observed coarsening of the instability wavelength. We finally compare the experimental predictions for the flows at early and late time with experimental measurements, and discuss how changes in light intensity may be captured by a corresponding ``effective'' stresslet.}

\subsection*{{Duality between pushers and pullers}}

In experiments, the zigzag instability arises because the rotated pullers exert an active stress on the jet{'s} boundary, which is mathematically {identical} (up to a constant pressure, as we show in the joint paper \cite{joint_article}) {to} the stress exerted by pushers with the same stresslet magnitude but aligned with the jet. This is fundamentally a consequence of the quasi-2D geometry, where cells are all parallel to each other. {No such duality is present, for instance, between a radial distribution of pullers and an axial distribution of pushers in a $3$D cylindrical jet}. Indeed, the flow around a 3D puller/pusher is axisymmetric about the swimmer's axis and not invariant under rotation.

\end{document}